\documentclass[prd,twocolumn,showpacs,amsmath,amssymb,superscriptaddress]{revtex4-1}
\pdfoutput=1

\usepackage[latin1]{inputenc}
\usepackage[dvipsnames]{xcolor}
\usepackage{mathtools}
\usepackage{subfigure}
\usepackage{hyperref}

\DeclareMathSymbol{\shortminus}{\mathbin}{AMSa}{"39}

\begin{document}
\author{Jos\'e Eliel Camargo-Molina}
\email{eliel.camargo-molina@physics.uu.se}
\affiliation{Department of Physics, Imperial College London, Blackett Laboratory, London, SW7 2AZ, UK}
\affiliation{Department of Physics and Astronomy, Uppsala University, Box 516, SE-751 20 Uppsala, Sweden}

\author{Arttu Rajantie}
\email{a.rajantie@imperial.ac.uk}
\affiliation{Department of Physics, Imperial College London, Blackett Laboratory, London, SW7 2AZ, UK}
\title{Phase transitions in de Sitter: The stochastic formalism}

\begin{abstract}
The stochastic spectral expansion method offers a simple framework for calculations in de Sitter spacetimes. We show
how to extend its reach to metastable vacuum states, both in the case when the potential is bounded from below, and when it is unbounded from below and therefore no stable vacuum state exists.
In both cases, the decay rate of the metastable vacuum is given by the lowest non-zero eigenvalue associated to the Fokker-Planck equation. We show how the corresponding eigenfunction determines the field probability distribution which can be used to compute correlation functions and other observables in the metastable vacuum state.
\noindent 
\end{abstract}

\maketitle

\section{introduction}
Recently, there has been significant interest in understanding vacuum instability of scalar field theories in the early Universe~\cite{Markkanen:2018pdo},
largely motivated by the observation that the electroweak vacuum state appears to be unstable in the Standard Model for the measured values of its parameters~\cite{Degrassi:2012ry,Buttazzo:2013uya}. 
In order for the Universe to exist in the electroweak vacuum today, its decay rate must have been sufficiently low throughout the cosmological history such that no vacuum decay events have taken place anywhere in our past lightcone. This requirement can place significant constraint on cosmological scenarios~\cite{Markkanen:2018pdo} and, for example, on the value of the non-minimal curvature coupling of the Higgs field~\cite{Herranen:2014cua,Herranen:2015ima}.

For vacuum decay during inflation, one can simplify calculations considerably by approximating the inflationary metric with the de Sitter spacetime. In that case, there are two main approaches for calculating the vacuum decay rate: the instanton method~\cite{Coleman:1980aw,Hawking:1981fz}, and the stochastic Starobinsky-Yokoyama method~\cite{Starobinsky:1986fx,Starobinsky:1994bd}. In this paper, we will focus on the latter, and show how it can be used to determine non-perturbatively both the vacuum decay rate and also, to a certain extent, the properties of the metastable state.

The stochastic formalism~\cite{Starobinsky:1994bd} provides a powerful framework to deal with light scalar fields in de Sitter space. It uses a stochastic Langevin equation to describe the dynamics of the long-wavelength field modes, with a noise term that arises from the short-wavelength quantum modes. The method has been used by many authors to calculate the rate of vacuum decay~\cite{Espinosa:2007qp,Hook:2014uia,Kearney:2015vba,Espinosa:2015qea,East:2016anr,Noorbala:2018zlv,Fumagalli:2019ohr}, using different prescriptions which give slightly different results. Although they agree on the stochastic equations themselves, there has been no consensus on how the decay rate should be defined in the context of the stochastic theory.

The spectral expansion method~\cite{risken1989fpe,Starobinsky:1994bd,Motohashi:2012bb,Markkanen:2019kpv,Markkanen:2020bfc} gives a powerful technique for solving the stochastic system in terms of a the eigenvalues and eigenfunctions of a Schr\"odinger-like equation. In this paper we investigate vacuum decay using this approach. It is known that the vacuum decay rate is given by the lowest non-zero eigenvalue~\cite{risken1989fpe}, which can be obtained to very high precision by solving the eigenvalue equation. As we will show, this result applies both when the potential is bounded from below and when it is unbounded, in which case no stable vacuum state exists. 

When considering cosmological scenarios in which the observer itself is in the metastable vacuum state, as appears to be the case for the electroweak vacuum, it is also important to be able to compute observables such as correlation functions in that state. This requires knowledge of the field probability distribution in the false vacuum state. We show that for potentials that are unbounded from below, this can be obtained unambiguously from the eigenfunctions of the eigenvalue equation associated to the Fokker-Planck equation. We also find a function that can be given the same interpretation when the potential is bounded from below.

\section{The Stochastic Formalism}\label{Sec:Stochastic}

Consider a scalar field $\phi$ in de Sitter space with potential $V(\phi)$ such that the field is light ($V^{\prime \prime}(\phi) < H^2$) and its contribution to the total energy is small ($V < 3 H^2 M_P^2$). Under those assumptions, the stochastic approach starts by treating the field as mainly long-wave modes (which can be thought as $\phi$ averaged over a constant volume slightly larger than a Hubble volume) and short-wave quantum modes $\xi$ that are modelled as a white noise term with the appropriate correlation properties. With this treatment, it is possible to write a Langevin equation satisfied by the long-wave modes~\cite{Starobinsky:1994bd} (which we call $\phi$ from now on), 
\begin{eqnarray}\label{eq:stochasticEq}
    \frac{d}{dt}\phi &=& -\frac{1}{3 H} V^\prime(\phi) + \xi(t) \\  &\mbox{with }& \langle \xi(t_1) \nonumber \xi(t_2) \rangle = \frac{H^3}{4 \pi^2}\delta(t_1 - t_2).
\end{eqnarray}
From this, it can be derived \cite{Starobinsky:1994bd} that the probability distribution for $\phi$, $P(t,\phi)$, satisfies the Fokker-Planck equation

\begin{equation}\label{eq:Fokker-Planck}
    \frac{\partial P(t,\phi)}{\partial t} = \frac{H^3}{4\pi^2}\left(
    \frac{1}{2}\frac{\partial^2}{\partial \phi^2} +v'\frac{\partial}{\partial \phi} + v'' \right)P(t,\phi),
\end{equation}
where we have introduced the re-scaled potential $v(\phi) = \frac{4\pi^2}{3 H^4} V(\phi)$. 

This equation admits linearly independent solutions of the form 
\begin{equation}
P_n(t,\phi) = e^{-v(\phi)} e^{-\Lambda_n t}\psi_n(\phi),
\end{equation}

where $\psi_n$ and $\Lambda_n$ are eigenfunctions and eigenvalues, respectively, of the differential equation

\begin{equation}\label{Eq:SchrEq}
   \left[-\frac{1}{2} \frac{\partial^2}{\partial \phi^2} + W(\phi)\right] \psi_n= \frac{4\pi^2 \Lambda_n}{H^3}  \psi_n, 
\end{equation}
with
\begin{equation}
    W(\phi) = \frac{1}{2} (v^\prime(\phi)^2 - v^{\prime \prime}(\phi)). 
\end{equation}
The eigenfunctions are chosen to be orthonormal,
\begin{equation}
    \int d\phi\,\psi_m(\phi)\psi_n(\phi)=\delta_{mn}.
\end{equation}
It is straightforward to check that the function
\begin{equation}\label{eq:groundstate}
    \psi_0(\phi) \propto e^{-v(\phi)},
\end{equation}
satisfies Eq.~(\ref{Eq:SchrEq}) with $\Lambda_0=0$ and if it satisfies the appropriate boundary conditions, it is therefore an eigenfunction with zero eigenvalue.

The time evolution of the probability distribution can therefore be expressed as
\begin{equation}\label{eq:pexpansion}
    P(t;\phi) = \psi_0(\phi) \sum_n a_n \psi_n(\phi) \, e^{-\Lambda_n (t - t_0)}
\end{equation}
where the coefficients $a_n$ can be determined from the initial conditions
as
\begin{equation}\label{eq:pcoeff}
    a_n=\int d\phi \frac{\psi_n(\phi)}{\psi_0(\phi)}P(t_0;\phi).
\end{equation}

It is useful to note that Eq.~(\ref{Eq:SchrEq}) has the form of the time-independent Schrodinger equation with the Hamiltonian
\begin{equation}
    {\cal H}=-\frac{1}{2} \frac{\partial^2}{\partial \phi^2} + W(\phi),
\end{equation}
where role of the potential is played by the function $W(\phi)$. Furthermore, this Hamiltonian is supersymmetric~\cite{SUSYQM} in the sense that it can be written as ${\cal H}=A^\dagger A$,
where 
\begin{equation}
A = \frac{1}{\sqrt{2}}\left(  \frac{d}{d\phi} + v^\prime(\phi) \right),\quad A^\dagger = \frac{1}{\sqrt{2}}\left(-\frac{d}{d\phi} + v^\prime(\phi)\right).
\end{equation}

This Hamiltonian has a superpartner
\begin{equation}
  \mathcal{\tilde{H}} = A A^\dagger =  -\frac{1}{2} \frac{\partial^2}{\partial \phi^2} + \tilde{W}(\phi),
\end{equation}
where
\begin{equation}\label{Eq:Ws}
\tilde{W}(\phi) = \frac{1}{2} (v^\prime(\phi)^2 + v^{\prime \prime}(\phi)),
\end{equation}
with the property that if $\psi$ is an eigenfunction of ${\cal H}$ with eigenvalue $\Lambda$, then
$\tilde\psi=A\psi$ satisfies
\begin{equation}
    \tilde{\cal H}\tilde\psi=AA^\dagger A\psi=A{\cal H}\psi=\Lambda A\psi=\Lambda\tilde\psi,
    \end{equation}
and is therefore an eigenfunction of ${\cal H}$ with the same eigenvalue $\Lambda$. The exception to this is the function $\psi_0$ defined in Eq.~(\ref{eq:groundstate}),
which satisfies $A\psi_0=0$, and therefore maps to zero under the supersymmetry transformation.

Therefore, as a consequence of supersymmetry, the spectra of ${\cal H}$ and $\tilde{\cal H}$ are identical, apart from possible zero eigenvalues, and the eigenfunctions are related by the mapping
\begin{equation}\label{eq:SUSYTrans}
    \tilde\psi_n=A\psi_n,\quad \psi_n=\frac{1}{\Lambda_n}A^\dagger\tilde\psi_n.
\end{equation}
The form of ${\cal H}$ also implies that the eigenvalues $\Lambda_n$ are non-negative.

It is also worth noting that the $\tilde{W}$ defined in Eq.~(\ref{Eq:Ws}) can be written
as
\begin{equation}
    \label{eq:Ws2}
    \tilde{W}(\phi) = \frac{1}{2} (\tilde{v}^\prime(\phi)^2 - \tilde{v}^{\prime \prime}(\phi)),
\end{equation}
with $\tilde{v}(\phi)=-v(\phi)$. Therefore the supersymmetry transformation can be interpreted as flipping the sign of the potential $V(\phi)$.

\section{Vacuum decay for bounded potentials}
\label{Sec:VacuumDecay}

Let us now assume that the potential is bounded from below, and grows sufficiently fast when $\phi\rightarrow\pm \infty$, so that the function
\begin{equation}\label{eq:Peq}
    P_{\rm eq}(\phi)=\psi_0(\phi)^2\propto e^{-2v(\phi)}
\end{equation}
is normalisable. In that case,
we can see from Eq.~(\ref{eq:pexpansion}) that it
is a time-independent solution of the Fokker-Planck equation (\ref{eq:Fokker-Planck}) and therefore corresponds to the equilibrium state of the system.

Let us also assume that the lowest non-zero eigenvalue is much smaller than the others, $\Lambda_1\ll \Lambda_2$.
This is the case when the potential $V(\phi)$ has a local minimum separated from the global minimum by a sufficiently high potential barrier. In classical field theory we would then identify the local minimum as a false vacuum state, and the global minimum as the true vacuum. Our aim now is to find the corresponding interpretation in the stochastic theory, in which states correspond to probability distributions rather than single field values.

Assuming this hierarchy of eigenvalues, the probability distribution at asymptotically late times is given by
\begin{equation}\label{eq:latetimes}
 P(t; \phi) = P_{\rm eq}(\phi) + c_0 \psi_0(\phi) \psi_1(\phi) e^{-\Lambda_1 t} + O\left(e^{-\Lambda_2 t}\right), 
\end{equation}
where again $c_0$ is a constant that can be determined from initial conditions. 

Alternative to the paragraph above: To interpret this, assume that in the false vacuum state the field has a probability distribution $P_1(\phi)$, which we would like to determine. If the system is initially (at time $t_0$) in this metastable state, we expect it to have a decreasing probability $p_1(t)=\exp(-\Gamma(t-t_0))$ of being still in the false vacuum state $P_1(\phi)$. Here $\Gamma$ is the false vacuum decay rate. Correspondingly, the system has probability $(1-p_1)$ of being in the true vacuum state $P_{\rm eq}(\phi)$.

The field probability distribution in such a mixed state is
\begin{eqnarray}\label{eq:twostates}
 P(t; \phi) &=&  (1 - p_1(t)) P_{\rm eq}(\phi) + p_1(t) P_1(\phi) \\ \nonumber
&=&P_{\rm eq}(\phi) + p_1(t) (P_1(\phi)-P_{\rm eq}(\phi)).
\end{eqnarray}
Comparing this with Eq.~(\ref{eq:latetimes}), we can see that this interpretation requires that $\Gamma=\Lambda_1$, i.e., the false vacuum decay rate is given by the lowest non-zero eigenvalue.

Furthermore, this identification suggests that we should be able to find the false vacuum probability distribution $P_1(\phi)$ by assuming the mixed state \eqref{eq:latetimes}, and following it back to the time $t_0$ when the system was fully in the false vacuum state. This gives
\begin{equation}
    \label{eq:P1}
    P_1(\phi)=P(t_0;\phi)=P_{\rm eq}(\phi)+\bar{c}_0\psi_0(\phi)\psi_1(\phi),
\end{equation}
where $\bar{c}_0=c_0\exp(-\Lambda_1 t_0)$.
This is a correctly normalised probability distribution because the eigenvalues are orthogonal,
\begin{eqnarray}
    \int d\phi \, P_1(\phi)&=&\int  d\phi \, P_{\rm eq}(\phi)+\bar{c}_0\int d\phi \, \psi_0(\phi)\psi_1(\phi)
    \nonumber\\
    &=&\int  d\phi \, P_{\rm eq}(\phi)=1.
\end{eqnarray}
However, it depends on the value that is chosen for the coefficient $\bar{c}_0$ or, equivalently, how far back in time one follows the evolution. In order for $P_1$ to be a well-defined probability distribution, it needs to be non-negative, and this means that the coefficient $\bar{c}_0$ has to be in the range
\begin{equation}
\label{eq:cbarrange}
-\frac{1}{\max \frac{\psi_1}{\psi_0}} \le \bar{c}_0\le 
    -\frac{1}{\min \frac{\psi_1}{\psi_0}} .
\end{equation}
Note that $\psi_0$ can be chosen to be positive, and $\psi_1$ has one zero. Therefore $\max\frac{\psi_1}{\psi_0}>0$ $\min\frac{\psi_1}{\psi_0}<0$.

This suggests that the two possible optimal choices for $\bar{c}_0$ are the two extremes of the range \eqref{eq:cbarrange}. In practice, the ratio $\psi_1/\psi_0$ is often a monotonic function and can be chosen to be an increasing function, in which case these two choices correspond to the limits of $\phi\rightarrow \infty$ and $\phi\rightarrow -\infty$, respectively. If the false minimum is, say, to the right of the true minimum, the appropriate choice for $P_1$ is the one in which the probability distribution is localised towards the positive values of $\phi$, which means that $\bar{c}_0>0$, and hence
\begin{equation}
    \bar{c}_0=-\frac{1}{\max \frac{\psi_1}{\psi_0}}=-\lim_{\phi\rightarrow\infty} \frac{\psi_0(\phi)}{\psi_1(\phi)}.
\end{equation}

In summary, when the potential is bounded from below, the decay rate of the false vacuum state is given by the lowest non-zero eigenvalue, $\Gamma=\Lambda_1$, and the field probability distribution in this false vacuum state can be written as
\begin{equation}
    P_1(\phi)=\psi_0(\phi)\left[\psi_0(\phi)-\left(\lim_{\phi'\rightarrow\infty} \frac{\psi_0(\phi')}{\psi_1(\phi')}\right)\psi_1(\phi)
    \right].
\end{equation}

\section{Vacuum decay for Unbounded potentials}

Interestingly, the stochastic formalism can be also applied to potentials that are unbounded from below. Such
 potentials are interesting for many reasons; the Standard Model potential at very high energies can be described by a negative quartic potential, and more generally, effective field theories with unknown high-energy origin can be described with unbounded potentials without necessarily jeopardizing their physical applicability.

Let us assume that we have a potential $V(\phi)$ that is finite everywhere but is not bounded from below, so that it approaches $-\infty$ as either $\phi\rightarrow\infty$ or $\phi\rightarrow -\infty$, or both.
In that case the function $\psi_0(\phi)$ defined by Eq.~(\ref{eq:groundstate}) still satisfies the eigenvalue equation (\ref{Eq:SchrEq}) with $\Lambda_0=0$, but it does not satisfy the correct boundary conditions and it is therefore not a valid eigenfunction. Correspondingly,
$P_{\rm eq}(\phi)$ defined by Eq.~(\ref{eq:Peq}) is not normalisable, and therefore does not give an equilibrium probability distribution.

Instead, the lowest eigenvalue, which we denote by $\Lambda_1$ is positive, and the corresponding eigenfunction $\psi_1(\phi)$ allows us to define the probability distribution
\begin{equation}\label{eq:P1ub}
    P_1(\phi) = \frac{1}{N} \psi_0(\phi) \psi_1(\phi),
\end{equation}
where the normalisation constant is
\begin{equation}\label{eq:Ndef}
    N=\int d\phi\, \psi_0(\phi)\psi_1(\phi).
\end{equation}
Note that because $\psi_0$ is not an eigenfunction, it is not orthogonal to $\psi_1$, and therefore $N\neq 0$.
Because $\psi_1$ is the lowest eigenfunction, it has no zeros, and therefore $P_1(\phi)$ is a non-negative function.

Of course, $P_1(\phi)$ is not an equilibrium probability distribution as such. If the field has initially, at time $t_0$, the probability distribution $P_1(\phi)$, then it follows from Eqs.~(\ref{eq:pexpansion}) and (\ref{eq:pcoeff}) that at a later time $t$, it has probability distribution
\begin{equation}\label{eq:quasieq}
    P(t;\phi)=e^{-\Lambda_1(t-t_0)}P_1(\phi).
\end{equation}
This shows concretely that probability is not conserved in the case of an unbounded potential. This is because there is a non-zero probability per unit time $\Gamma=\Lambda_1>0$ that the field rolls down the potential and reaches infinity. Therefore, just like in the case of the bounded potential, $\Lambda_1$ gives the vacuum decay rate.

If we consider an observer whose existence requires the field $\phi$ to have a finite value, and which gets destroyed if the field ever reaches infinity, then the observer will only ever observe the conditional probability distribution which assumes that the field is finite. At any time $t$, this is given by
\begin{equation}
    P\Bigr(t;\phi\Big||\phi|<\infty\Bigr)
    =\frac{P(t;\phi)}{\int_{-\infty}^\infty d\phi' P(t;\phi')}
    =P_1(\phi).
\end{equation}
Therefore, the observer actually observes the field in an time-independent probability distribution $P_1(\phi)$.

For this construction to work, the normalisation constant $N$ defined in Eq.~(\ref{eq:Ndef}) must be finite. This is not obvious because $\psi_0(\phi)$ diverges at infinity. 
We can use the supersymmetry transformation to investigate this. If we assume that $\lim_{\phi\rightarrow\pm\infty}v(\phi)=-\infty$, the superpartner $\tilde{v}(\phi)=-v(\phi)$ is bounded from below. Its lowest eigenfunction, with zero eigenvalue, is
\begin{equation}
    \tilde\psi_0(\phi)\propto e^{-\tilde{v}(\phi)}=e^{v(\phi)}.
\end{equation}
We can also use the perturbative techniques from Ref.~\cite{Starobinsky:1994bd} to find the asymptotic behaviour of the next eigenfunction $\tilde\psi_1(\phi)$ at large field values,
\begin{eqnarray}\label{Eq:StaroYokoPsi1}
\nonumber
    \tilde\psi_1(\phi) &\propto & \tilde{\psi}_0(\phi) \\ 
&&
    -2 \epsilon_1\tilde{\psi}_0(\phi)
    \int_\phi^\infty\, d\phi'\, 
    \int_{\phi}^{\phi'}\, d\phi''\,
    e^{2\tilde{v}(\phi'')-2\tilde{v}(\phi')}
    ,
\end{eqnarray}
where $\epsilon_1 = 4 \pi^2 \Lambda_1/H^3 $ is the perturbative expansion parameter.
Applying the inverse supersymmetry transformation (\ref{eq:SUSYTrans}), we find the perturbative expression for the lowest eigenstate in the original unstable theory,
\begin{equation}
\psi_1(\phi)=\frac{1}{\Lambda_1}A^\dagger\tilde{\psi}_1(\phi)
\propto e^{-v(\phi)}\int_\phi^\infty d\phi' e^{2v(\phi')}.
\end{equation}
Note that, in fact, this is the same as $\Phi_{\rm st}^{(1)}$ defined in Eq.~(52) of \cite{Starobinsky:1994bd}.

If we assume that $v(\phi)\sim -g\phi^\alpha$, $\alpha>0$, as $\phi\rightarrow\infty$, then 
\begin{equation}
    \psi_1(\phi)\sim \Gamma\left(\frac{1}{\alpha},\phi^\alpha\right) \sim \phi^{1-\alpha}e^{-g\phi^\alpha},
\end{equation}
where $\Gamma(s,x)$ is the incomplete gamma function.
The probability distribution $P_1(\phi)$ then behaves asymptotically as
\begin{equation}\label{eq:asymptP1}
    P_1(\phi)\sim \phi^{1-\alpha},
\end{equation}
and is normalisable if $\alpha>2$. Therefore the construction works for unstable potentials that are steeper than harmonic.

As a consistency check, we can also see that the time-dependent probability distribution $P(t;\phi)$ satisfies the continuity equation
\begin{equation}
    \frac{\partial P}{\partial t}=\frac{\partial}{\partial \phi}J,
\end{equation}
where 
\begin{equation}
    J(\phi)=\frac{H^3}{4\pi^2}\left(\frac{1}{2}\frac{\partial P}{\partial \phi}+v'P\right).
\end{equation}
Integrating over $\phi$, Eq.~(\ref{eq:quasieq}) implies
\begin{equation}
    -\Lambda_1=\lim_{\phi\rightarrow\infty}\left(J(\phi)-J(-\phi)\right)=
    \lim_{\phi\rightarrow\infty}\frac{H^3}{2\pi^2}v'(\phi)P_1(\phi),
\end{equation}
where we have assumed a symmetric potential, $v(-\phi)=v(\phi)$, for simplicity. From this we can see that
we must have
\begin{equation}\label{equ:P1asymp}
    P_1(\phi)\sim \frac{2\pi^2\Lambda_1}{H^3}\frac{1}{|v'(\phi)|},
\end{equation}
which is consistent with Eq.~(\ref{eq:asymptP1}). In the case of non-symmetric potentials, while Eq.~\eqref{equ:P1asymp} would have a different constant prefactor, it would still be proportional to $1/|v'(\phi)|$.

In summary, for an unbounded potential, the vacuum decay rate is give by $\Gamma=\Lambda_1$, just like for bounded potentials, and the false vacuum ``equilibrium'' probability distribution is given by Eq.~(\ref{eq:P1ub}).

\section{Numerical Results}
\label{Sec:Numerics}
In this section we take our discussion above and apply it to a concrete case. We start by considering the bounded scalar potential 
\begin{equation}
   V^+(\phi) = \frac{3 H^4}{4 \pi^2} v^+ = \mu^3 \phi - \frac{1}{2}\bar{m}^2 \phi^2 + \lambda \phi^4,
\end{equation}
where $\lambda>0$,
and the unbounded potential
$V^-(\phi)=-V^+(\phi)$, which is also the superpartner of $V^+$. In the following we will use superscripts $+$ and $-$ to indicate whether the quantity relates to the bounded or unbounded from below cases respectively.

\begin{figure}[tbh!]%
    {{\includegraphics[width=0.47\textwidth]{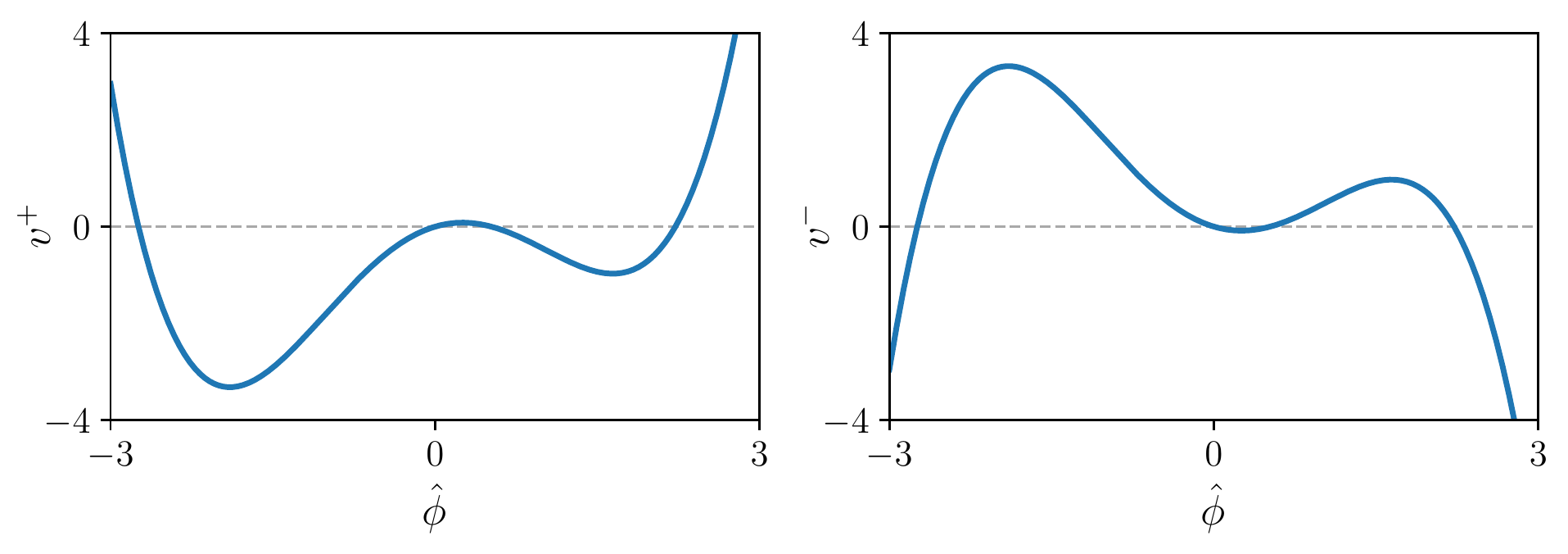} }}
\caption{The shapes of $v^+$ (left) and $v^-$ (right) for $\bar{\alpha} = 0.8$ and $\beta=0.1$.
}
\label{Fig:vs}%
\end{figure}

As help for numerical calculations, we will cast the re-scaled potential $v^+$ as a function of dimensionless parameters $\bar{\alpha} = \bar{m}^2 / \lambda^{\frac{1}{2}} H^2$, $\beta = \mu^3 / \lambda^{\frac{1}{4}} H^3$ and the dimensionless scalar field $\hat{\phi} = \frac{\lambda^{\frac{1}{4}} \Omega}{H} \phi$ where $\Omega = 1 + \sqrt{\bar{\alpha}}+\beta$. This results in the dimensionless potential
\begin{equation}
\frac{3 \Omega}{\pi^2} v^+ = 4\beta \hat{\phi} + 2 \bar{\alpha} \hat{\phi}^2 + \hat{\phi}^4.
\end{equation}
We show $v^+$ and $v^- = -v$ in Fig.~\ref{Fig:vs} (for $\bar{\alpha} = 0.8$ and $\beta=0.1$) to illustrate that one is a bounded potential with true and false vacua and the other an unbounded potential with a minimum around the origin.
\begin{figure}[tbh!]%
    {{\includegraphics[width = 0.47\textwidth]{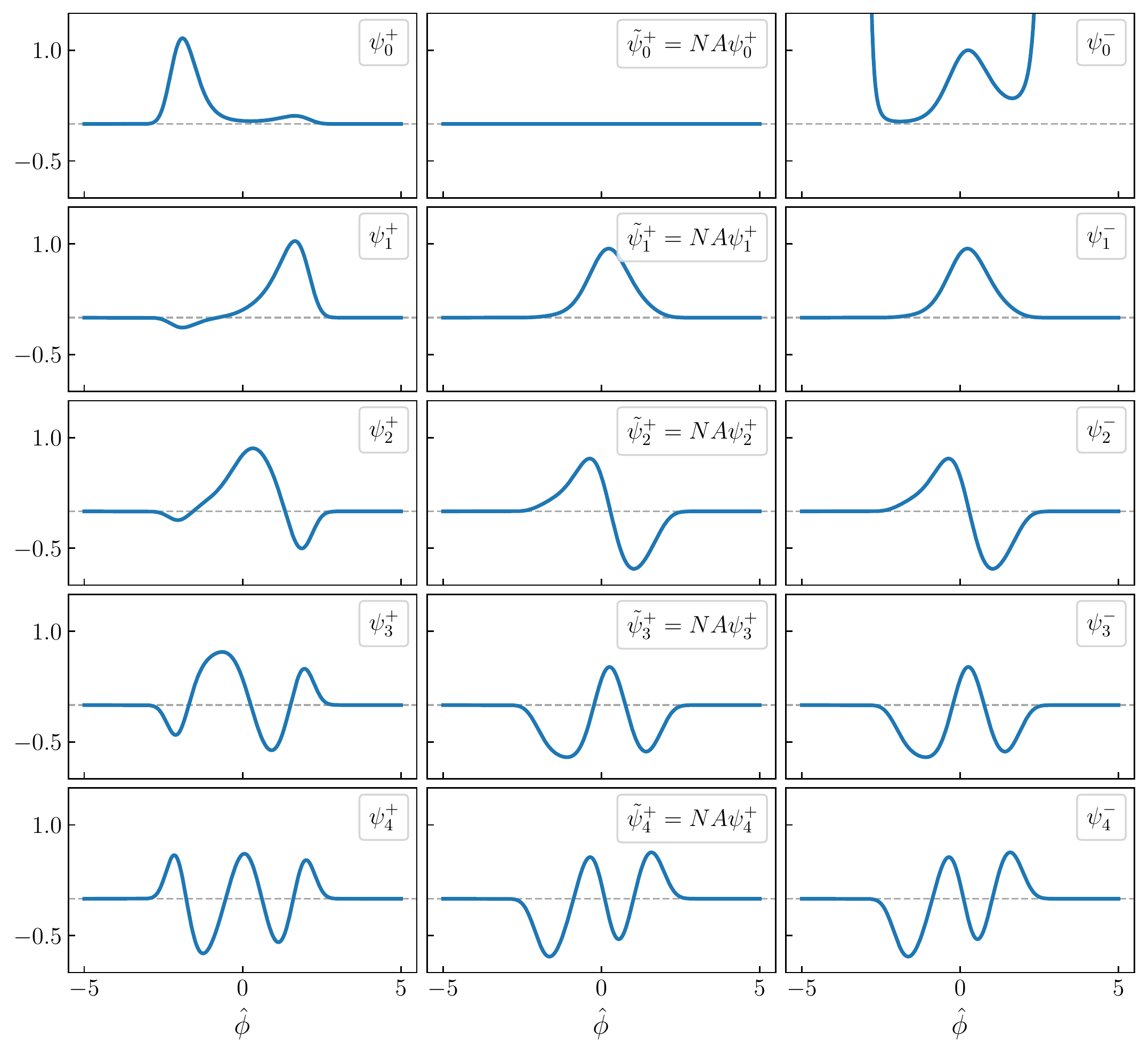} }}
    \caption{Numerical results for the SUSY transformation of the eigenfunctions of \eqref{Eq:SchrEq} for $\bar{\alpha} = 0.8$ and $\beta = 0.1$. Left: Eigenfunctions of the bounded potential, center: The SUSY-transformed eigenfunction, right: Eigenfunctions of the unbounded potential. All eigenfunctions have been re-scaled to help visualization.}
    \label{Fig:susyqm}%
\end{figure}

\begin{figure}[tbh!]%
    {{\includegraphics[width = 0.47\textwidth]{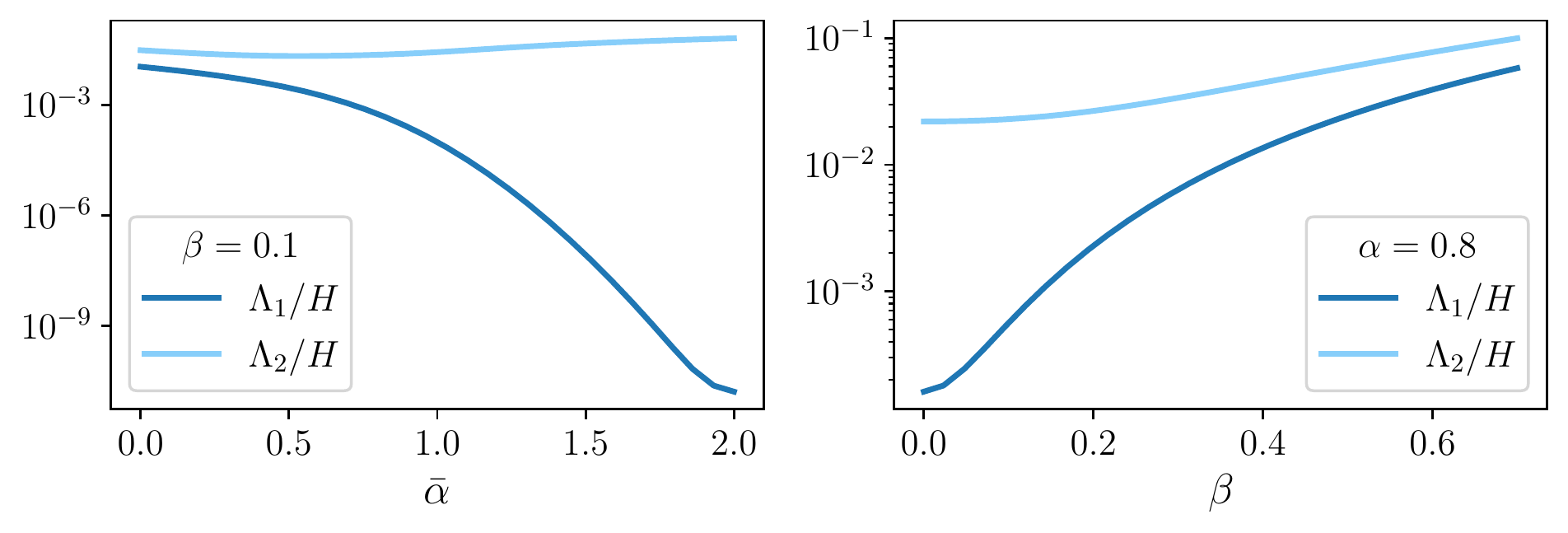} }}
    \caption{Change of $\Lambda_1$ and $\Lambda_2$ as function of the potential parameters around the point $\bar{\alpha} = 0.8$, $\beta=0.1$. Note that this figure is identical in both bounded and unbounded from below cases.}
    \label{Fig:lambdas}%
\end{figure}

The eigenvalue equation~\eqref{Eq:SchrEq} can be easily solved numerically to get the eigenvalues $\Lambda^+_i$ ($\Lambda^-_i$) and the eigenfunctions $\psi^+_i$ ($\psi^-_i$) of $\mathcal{H^+}$ ($\tilde{\mathcal{H}}^+ = \mathcal{H}^-$), the Hamiltonian corresponding to $v^+$ ($v^-$). In this work we chose to use Mathematica's \texttt{NDEigensystem}. In Fig.~\ref{Fig:lambdas} we show $\Lambda_1$ and $\Lambda_2$ as a function of $\bar{\alpha}$ and $\beta$. For larger values of $\bar{\alpha}$ the potential barrier grows and so does the hierarchy, making $\Lambda_1 \ll \Lambda_2$ as expected. For larger $\beta$ the hierarchy decreases as the relative depth of the true vacuum increases and the barrier height lowers. For large enough $\beta$ the false vacuum dissappears. 

As discussed in Sec.~\ref{Sec:Stochastic}, supersymmetry implies that the eigenvalues of the two Hamiltonians should be equal, and the eigenfunctions should be related by the supersymmetry transformation (\ref{eq:SUSYTrans}).
We show this in in Fig.~\ref{Fig:susyqm}. It is interesting to note that the SUSY transformation, as expected, does not work for the ground state $\psi_0^+ \propto e^{-v^+}$. As we discussed, $\psi_0^+$ corresponds to a vanishing eigenvalue and gets annihilated by the SUSY transformation. For clarity, we show the function $\psi_0^- \propto e^{-v^-}$, which diverges at $z \rightarrow \infty$ and thus not really part of $\mathcal{H}^-$'s spectrum.    

By choosing values for $\bar{\alpha}$ and $\beta$, we can see how different scenarios lead to qualitatively different probability distributions for the scalar field. 

\begin{figure}[tbh!]%
    \includegraphics[width = 0.5\textwidth]{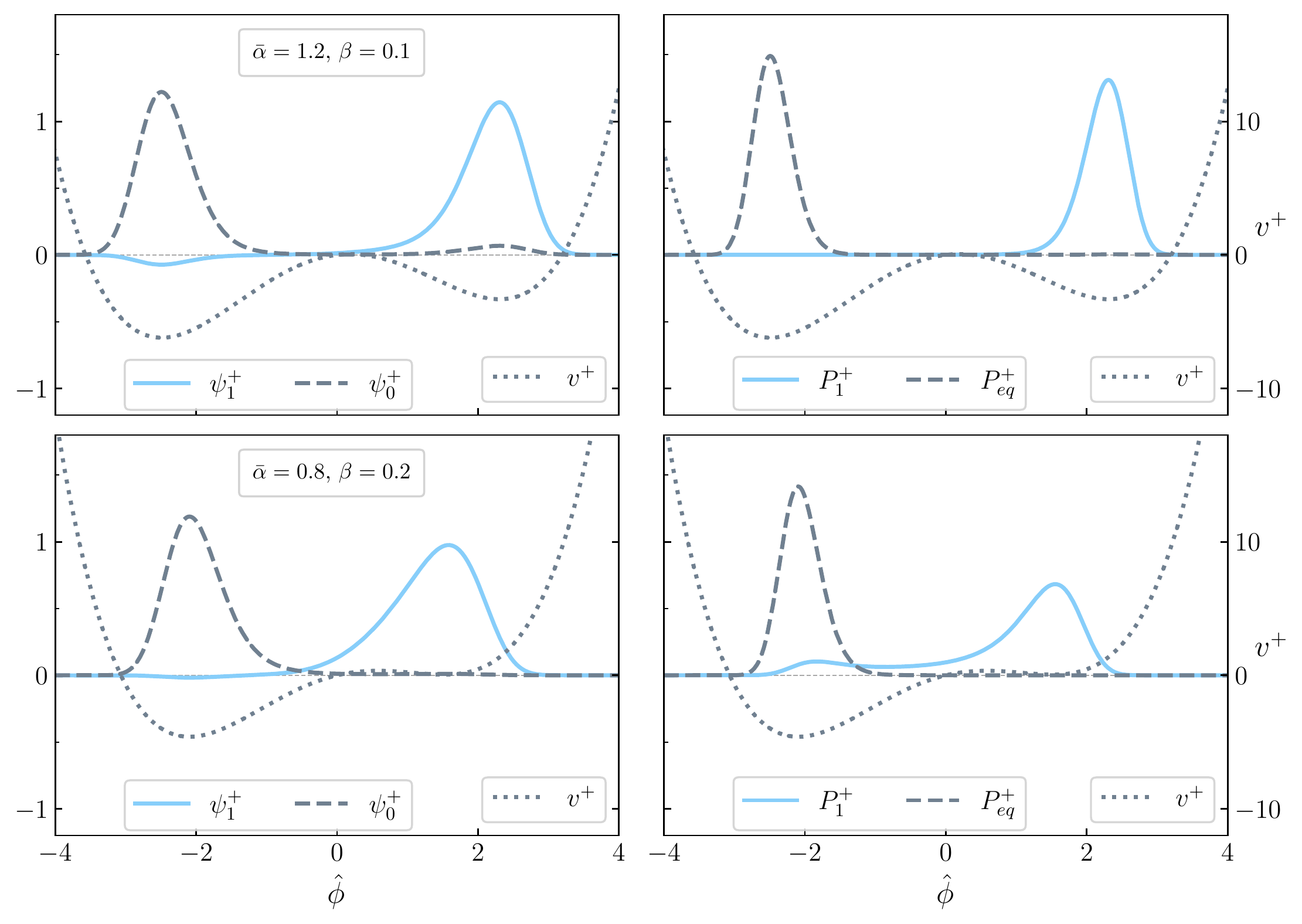} 
\caption{Comparison between $\psi^+_0$, $\psi^+_1$ (left) and $P_{\rm eq}$, $P^+_1$ (right) for two sets of parameter values. The corresponding potentials are shown for illustration.}
\label{Fig:comparissonP1Plus}%
\end{figure}

Considering the bounded case $V^+(\phi)$ first, the first two eigenfunctions $\psi_0^+$ and $\psi_1^+$ are shown in Fig.~\ref{Fig:comparissonP1Plus}, as well as the functions $P_1^+$ given by Eq.~(\ref{eq:P1}) and $P^+_{\rm eq}$ given by Eq.~(\ref{eq:Peq}). In Sec.~\ref{Sec:VacuumDecay}, it was argued that $P_1^+$ can be interpreted as the probability distribution of the field $\phi$ in the metastable vacuum. For $\bar{\alpha}=1.2$, when the barrier between the false and true vacua is high, it is localised around the false vacuum, in agreement with this interpretation. For $\bar{\alpha}=0.8$, the barrier is lower, and in that case the probability distribution $P_1^+$ extends to the true vacuum side.

For the unbounded case $V^-(\phi)$, the field probability distribution in the metastable state $P_1^-$ is given by Eq.~(\ref{eq:P1ub}).

\begin{figure}[tbh!]%
    {{\includegraphics[width = 0.5\textwidth]{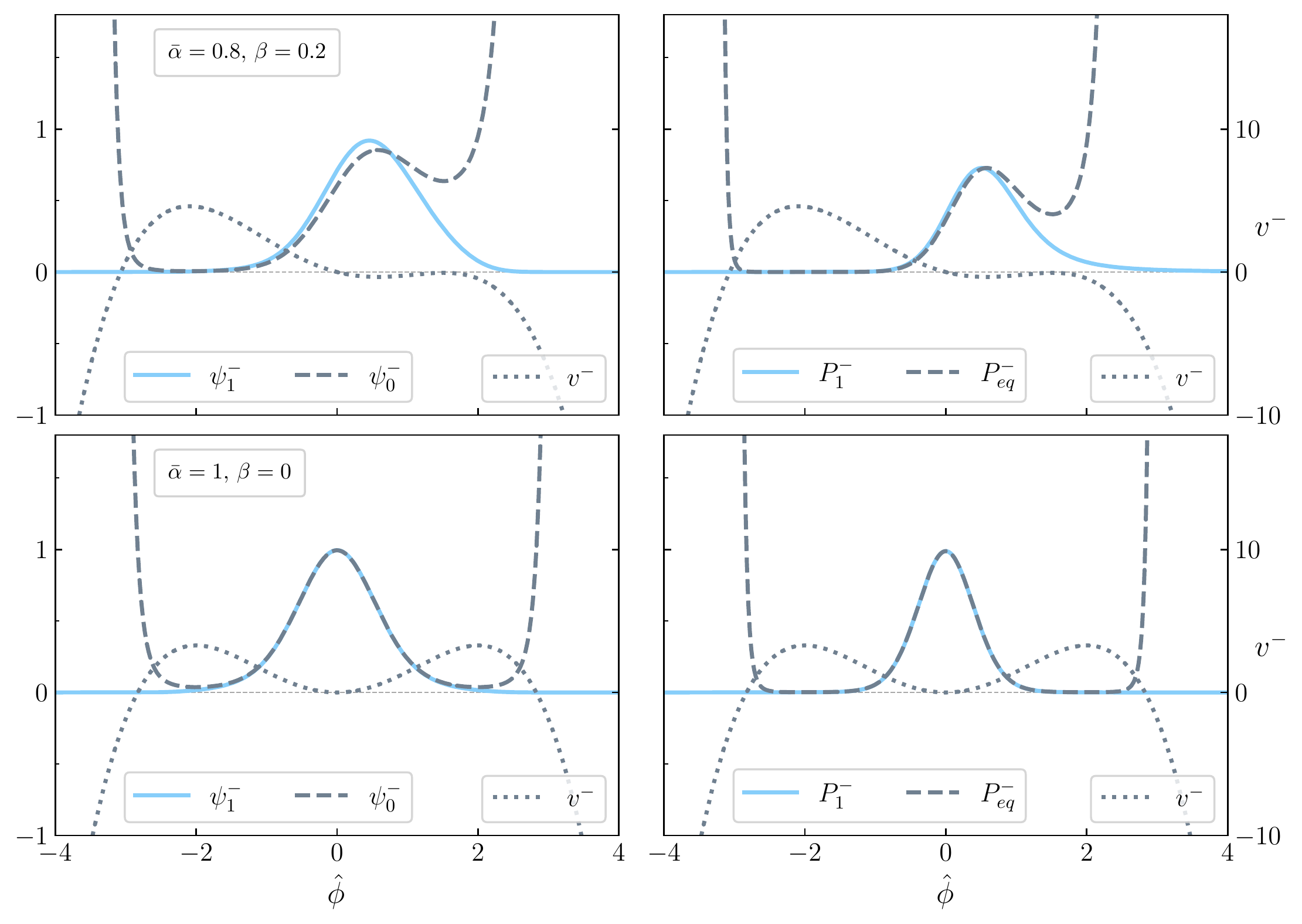} }}
\caption{Comparison between $\psi^-_0$, $\psi^-_1$ (left) and $P^-_{\rm eq}$, $P^-_1$ (right) for a non-symmetric (top) and symmetric (bottom) potentials. The corresponding potentials are shown for illustration.}
\label{Fig:comparissonP1}%
\end{figure}

This function is shown in Fig.~\ref{Fig:comparissonP1}, where we can see that it is indeed localised around the local minimum of the potential. For comparison, we are also showing the function $P^-_{\rm eq}$ defined also by Eq.~(\ref{eq:Peq}), which would be the equilibrium state for a bounded potential, but which is not normalisable in the unbounded case. We can see that near the local minimum, $P^-_1(\phi)\approx P^-_{\rm eq}(\phi)$, as one would expect, because if the lifetime of the metastable vacuum is sufficiently long, it should be almost indistinguishable from a stable vacuum state.


\begin{figure}[!htb]
\includegraphics[width = 0.47\textwidth]{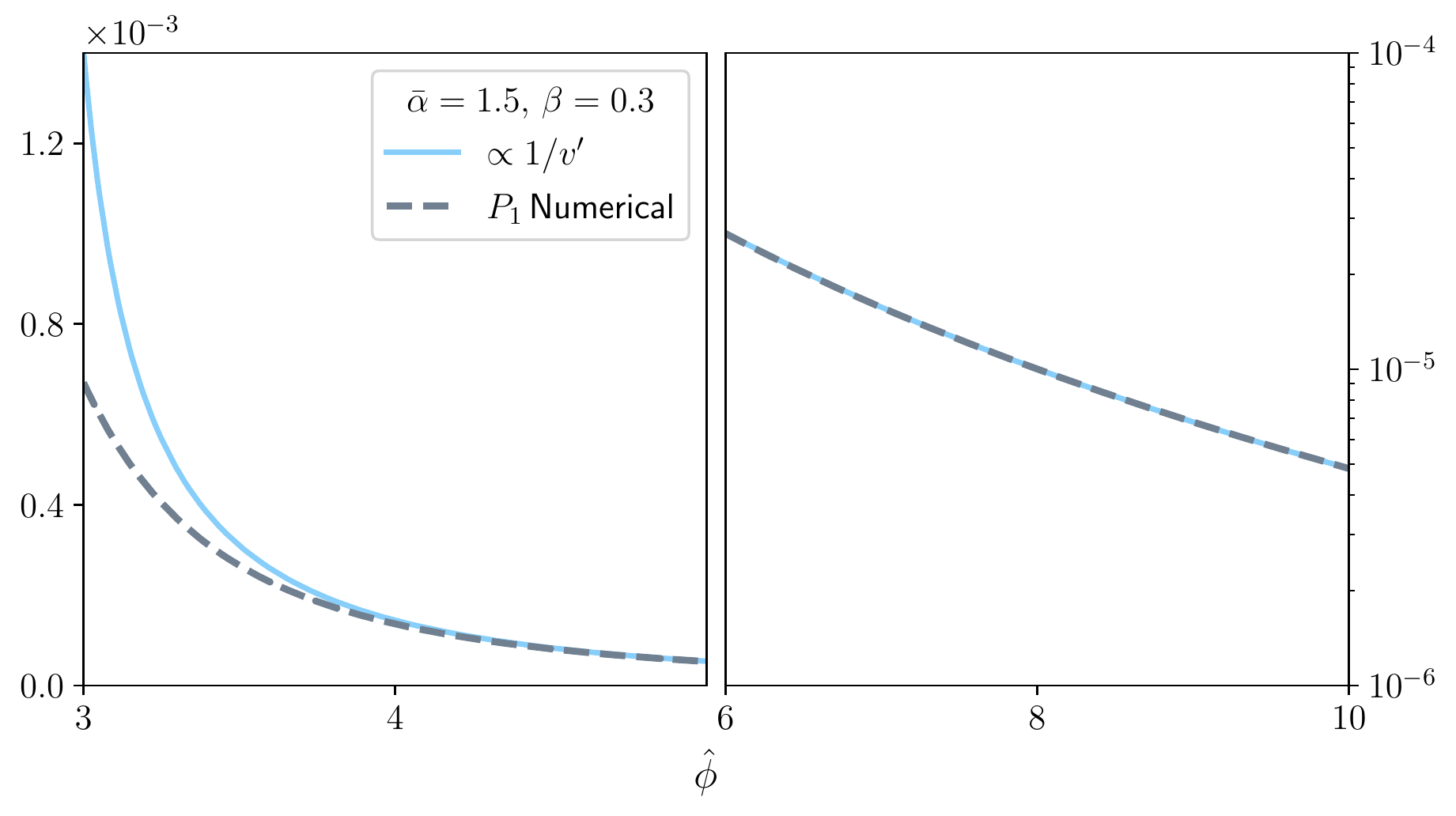}
\caption{The numerical probability density for large field values compared to Eq.~\eqref{equ:P1asymp}. On the left, for smaller field values, we show a linear scale while on the right we switch to logarithmic for larger field values. Both expressions agree very well from $\hat{\phi} \sim 4$ and continue to do so across several orders of magnitude.}
\label{fig:analyvsnum}
\end{figure}

We also explored the validity of the expected asymptotic behaviour, by comparing the full numerical calculation of $P_1^-$ to the expression in Eq.~\eqref{equ:P1asymp}. As the latter assumes a symmetric potential, there will be a relative constant factor when compared to the numerical results, though the crucial insight is that it will be proportional to $1/(v^-)'$ regardless. We checked this for a range of parameter combinations, showing very good agreement, and we show a typical example in Fig.~\ref{fig:analyvsnum}. Given that our analytical approximation is only valid for large field values, the agreement with the numerical results gets better for larger $\phi$.
\section{Discussion}

In this article, we have shown how the Starobinsky-Yokoyama stochastic approach can be used to describe vacuum decay in scalar field theories in de Sitter space, both in the case of potentials that are bounded and unbounded from below. In both cases, the decay rate per unit time of the metastable vacuum is given by the lowest non-zero eigenvalue of the eigenvalue equation associated to the Fokker-Planck equation, which is a known result from stochastic analysis~\cite{risken1989fpe}. 

We also showed that the corresponding eigenfunction determines the observables in the metastable vacuum state. In the case of an unbounded potential, the relation is straightforward and unambiguous. In bounded potentials, a probability distribution cannot be uniquely associated with the metastable vacuum state, but by following the time evolution backwards as far as possible, we determined a function that can be given that interpretation. These probability distributions are useful for computing predictions for observables that would be measured by an observer in the metastable vacuum.

The formalism and methods presented in this article, facilitate phenomenological studies of phase transitions and vacuum decay during inflation.
In the case of the Standard Model, which famously exhibits an unbounded potential at higher energies \cite{Degrassi:2012ry}, 
the prediction of a single vacuum decay event in the past light cone would rule out the theory, and hence require some new physics Beyond the Standard Model, phase transitions during inflation imply a primordial gravitational wave signature that will be probed at future experiments \cite{Chialva:2010jt}. This paper sets the ground for such precision calculations. 

In this article we focused on vacuum decay purely within the stochastic approach. While this helps set a clear interpretation of vacuum decay in the stochastic formalism, a precision calculation needs to nevertheless put our results in the context of quantum field theory. As shown in Ref.~\cite{Camargo-Molina:2022nnn}, it is then possible to e.g. incorporate quantum corrections at the one-loop order.

\section*{acknowledgments}
A.R. and J.E.C. were supported by by Science and Technology Facilities Council (UK) grant ST/P000762/1, and A.R. also by grant ST/T000791/1. J.E.C was supported by the Carl Trygger Foundation through grant no. CTS 17:139. The authors would like to thank Mariana Carrillo Gonz\'alez for useful discussions.

\bibliographystyle{apsrev4-2}
\bibliography{Bib.bib}

\begin{thebibliography}{23}%
\makeatletter
\providecommand \@ifxundefined [1]{%
 \@ifx{#1\undefined}
}%
\providecommand \@ifnum [1]{%
 \ifnum #1\expandafter \@firstoftwo
 \else \expandafter \@secondoftwo
 \fi
}%
\providecommand \@ifx [1]{%
 \ifx #1\expandafter \@firstoftwo
 \else \expandafter \@secondoftwo
 \fi
}%
\providecommand \natexlab [1]{#1}%
\providecommand \enquote  [1]{``#1''}%
\providecommand \bibnamefont  [1]{#1}%
\providecommand \bibfnamefont [1]{#1}%
\providecommand \citenamefont [1]{#1}%
\providecommand \href@noop [0]{\@secondoftwo}%
\providecommand \href [0]{\begingroup \@sanitize@url \@href}%
\providecommand \@href[1]{\@@startlink{#1}\@@href}%
\providecommand \@@href[1]{\endgroup#1\@@endlink}%
\providecommand \@sanitize@url [0]{\catcode `\\12\catcode `\$12\catcode
  `\&12\catcode `\#12\catcode `\^12\catcode `\_12\catcode `\%12\relax}%
\providecommand \@@startlink[1]{}%
\providecommand \@@endlink[0]{}%
\providecommand \url  [0]{\begingroup\@sanitize@url \@url }%
\providecommand \@url [1]{\endgroup\@href {#1}{\urlprefix }}%
\providecommand \urlprefix  [0]{URL }%
\providecommand \Eprint [0]{\href }%
\providecommand \doibase [0]{https://doi.org/}%
\providecommand \selectlanguage [0]{\@gobble}%
\providecommand \bibinfo  [0]{\@secondoftwo}%
\providecommand \bibfield  [0]{\@secondoftwo}%
\providecommand \translation [1]{[#1]}%
\providecommand \BibitemOpen [0]{}%
\providecommand \bibitemStop [0]{}%
\providecommand \bibitemNoStop [0]{.\EOS\space}%
\providecommand \EOS [0]{\spacefactor3000\relax}%
\providecommand \BibitemShut  [1]{\csname bibitem#1\endcsname}%
\let\auto@bib@innerbib\@empty
\bibitem [{\citenamefont {Markkanen}\ \emph {et~al.}(2018)\citenamefont
  {Markkanen}, \citenamefont {Rajantie},\ and\ \citenamefont
  {Stopyra}}]{Markkanen:2018pdo}%
  \BibitemOpen
  \bibfield  {author} {\bibinfo {author} {\bibfnamefont {T.}~\bibnamefont
  {Markkanen}}, \bibinfo {author} {\bibfnamefont {A.}~\bibnamefont
  {Rajantie}},\ and\ \bibinfo {author} {\bibfnamefont {S.}~\bibnamefont
  {Stopyra}},\ }\href {https://doi.org/10.3389/fspas.2018.00040} {\bibfield
  {journal} {\bibinfo  {journal} {Front. Astron. Space Sci.}\ }\textbf
  {\bibinfo {volume} {5}},\ \bibinfo {pages} {40} (\bibinfo {year} {2018})},\
  \Eprint {https://arxiv.org/abs/1809.06923} {arXiv:1809.06923 [astro-ph.CO]}
  \BibitemShut {NoStop}%
\bibitem [{\citenamefont {Degrassi}\ \emph {et~al.}(2012)\citenamefont
  {Degrassi}, \citenamefont {Di~Vita}, \citenamefont {Elias-Miro},
  \citenamefont {Espinosa}, \citenamefont {Giudice}, \citenamefont {Isidori},\
  and\ \citenamefont {Strumia}}]{Degrassi:2012ry}%
  \BibitemOpen
  \bibfield  {author} {\bibinfo {author} {\bibfnamefont {G.}~\bibnamefont
  {Degrassi}}, \bibinfo {author} {\bibfnamefont {S.}~\bibnamefont {Di~Vita}},
  \bibinfo {author} {\bibfnamefont {J.}~\bibnamefont {Elias-Miro}}, \bibinfo
  {author} {\bibfnamefont {J.~R.}\ \bibnamefont {Espinosa}}, \bibinfo {author}
  {\bibfnamefont {G.~F.}\ \bibnamefont {Giudice}}, \bibinfo {author}
  {\bibfnamefont {G.}~\bibnamefont {Isidori}},\ and\ \bibinfo {author}
  {\bibfnamefont {A.}~\bibnamefont {Strumia}},\ }\href
  {https://doi.org/10.1007/JHEP08(2012)098} {\bibfield  {journal} {\bibinfo
  {journal} {JHEP}\ }\textbf {\bibinfo {volume} {08}},\ \bibinfo {pages}
  {098}},\ \Eprint {https://arxiv.org/abs/1205.6497} {arXiv:1205.6497 [hep-ph]}
  \BibitemShut {NoStop}%
\bibitem [{\citenamefont {Buttazzo}\ \emph {et~al.}(2013)\citenamefont
  {Buttazzo}, \citenamefont {Degrassi}, \citenamefont {Giardino}, \citenamefont
  {Giudice}, \citenamefont {Sala}, \citenamefont {Salvio},\ and\ \citenamefont
  {Strumia}}]{Buttazzo:2013uya}%
  \BibitemOpen
  \bibfield  {author} {\bibinfo {author} {\bibfnamefont {D.}~\bibnamefont
  {Buttazzo}}, \bibinfo {author} {\bibfnamefont {G.}~\bibnamefont {Degrassi}},
  \bibinfo {author} {\bibfnamefont {P.~P.}\ \bibnamefont {Giardino}}, \bibinfo
  {author} {\bibfnamefont {G.~F.}\ \bibnamefont {Giudice}}, \bibinfo {author}
  {\bibfnamefont {F.}~\bibnamefont {Sala}}, \bibinfo {author} {\bibfnamefont
  {A.}~\bibnamefont {Salvio}},\ and\ \bibinfo {author} {\bibfnamefont
  {A.}~\bibnamefont {Strumia}},\ }\href
  {https://doi.org/10.1007/JHEP12(2013)089} {\bibfield  {journal} {\bibinfo
  {journal} {JHEP}\ }\textbf {\bibinfo {volume} {12}},\ \bibinfo {pages}
  {089}},\ \Eprint {https://arxiv.org/abs/1307.3536} {arXiv:1307.3536 [hep-ph]}
  \BibitemShut {NoStop}%
\bibitem [{\citenamefont {Herranen}\ \emph {et~al.}(2014)\citenamefont
  {Herranen}, \citenamefont {Markkanen}, \citenamefont {Nurmi},\ and\
  \citenamefont {Rajantie}}]{Herranen:2014cua}%
  \BibitemOpen
  \bibfield  {author} {\bibinfo {author} {\bibfnamefont {M.}~\bibnamefont
  {Herranen}}, \bibinfo {author} {\bibfnamefont {T.}~\bibnamefont {Markkanen}},
  \bibinfo {author} {\bibfnamefont {S.}~\bibnamefont {Nurmi}},\ and\ \bibinfo
  {author} {\bibfnamefont {A.}~\bibnamefont {Rajantie}},\ }\href
  {https://doi.org/10.1103/PhysRevLett.113.211102} {\bibfield  {journal}
  {\bibinfo  {journal} {Phys. Rev. Lett.}\ }\textbf {\bibinfo {volume} {113}},\
  \bibinfo {pages} {211102} (\bibinfo {year} {2014})},\ \Eprint
  {https://arxiv.org/abs/1407.3141} {arXiv:1407.3141 [hep-ph]} \BibitemShut
  {NoStop}%
\bibitem [{\citenamefont {Herranen}\ \emph {et~al.}(2015)\citenamefont
  {Herranen}, \citenamefont {Markkanen}, \citenamefont {Nurmi},\ and\
  \citenamefont {Rajantie}}]{Herranen:2015ima}%
  \BibitemOpen
  \bibfield  {author} {\bibinfo {author} {\bibfnamefont {M.}~\bibnamefont
  {Herranen}}, \bibinfo {author} {\bibfnamefont {T.}~\bibnamefont {Markkanen}},
  \bibinfo {author} {\bibfnamefont {S.}~\bibnamefont {Nurmi}},\ and\ \bibinfo
  {author} {\bibfnamefont {A.}~\bibnamefont {Rajantie}},\ }\href
  {https://doi.org/10.1103/PhysRevLett.115.241301} {\bibfield  {journal}
  {\bibinfo  {journal} {Phys. Rev. Lett.}\ }\textbf {\bibinfo {volume} {115}},\
  \bibinfo {pages} {241301} (\bibinfo {year} {2015})},\ \Eprint
  {https://arxiv.org/abs/1506.04065} {arXiv:1506.04065 [hep-ph]} \BibitemShut
  {NoStop}%
\bibitem [{\citenamefont {Coleman}\ and\ \citenamefont
  {De~Luccia}(1980)}]{Coleman:1980aw}%
  \BibitemOpen
  \bibfield  {author} {\bibinfo {author} {\bibfnamefont {S.~R.}\ \bibnamefont
  {Coleman}}\ and\ \bibinfo {author} {\bibfnamefont {F.}~\bibnamefont
  {De~Luccia}},\ }\href {https://doi.org/10.1103/PhysRevD.21.3305} {\bibfield
  {journal} {\bibinfo  {journal} {Phys. Rev. D}\ }\textbf {\bibinfo {volume}
  {21}},\ \bibinfo {pages} {3305} (\bibinfo {year} {1980})}\BibitemShut
  {NoStop}%
\bibitem [{\citenamefont {Hawking}\ and\ \citenamefont
  {Moss}(1982)}]{Hawking:1981fz}%
  \BibitemOpen
  \bibfield  {author} {\bibinfo {author} {\bibfnamefont {S.~W.}\ \bibnamefont
  {Hawking}}\ and\ \bibinfo {author} {\bibfnamefont {I.~G.}\ \bibnamefont
  {Moss}},\ }\href {https://doi.org/10.1016/0370-2693(82)90946-7} {\bibfield
  {journal} {\bibinfo  {journal} {Phys. Lett. B}\ }\textbf {\bibinfo {volume}
  {110}},\ \bibinfo {pages} {35} (\bibinfo {year} {1982})}\BibitemShut
  {NoStop}%
\bibitem [{\citenamefont {Starobinsky}(1986)}]{Starobinsky:1986fx}%
  \BibitemOpen
  \bibfield  {author} {\bibinfo {author} {\bibfnamefont {A.~A.}\ \bibnamefont
  {Starobinsky}},\ }\href {https://doi.org/10.1007/3-540-16452-9_6} {\bibfield
  {journal} {\bibinfo  {journal} {Lect. Notes Phys.}\ }\textbf {\bibinfo
  {volume} {246}},\ \bibinfo {pages} {107} (\bibinfo {year}
  {1986})}\BibitemShut {NoStop}%
\bibitem [{\citenamefont {Starobinsky}\ and\ \citenamefont
  {Yokoyama}(1994)}]{Starobinsky:1994bd}%
  \BibitemOpen
  \bibfield  {author} {\bibinfo {author} {\bibfnamefont {A.~A.}\ \bibnamefont
  {Starobinsky}}\ and\ \bibinfo {author} {\bibfnamefont {J.}~\bibnamefont
  {Yokoyama}},\ }\href {https://doi.org/10.1103/PhysRevD.50.6357} {\bibfield
  {journal} {\bibinfo  {journal} {Phys. Rev. D}\ }\textbf {\bibinfo {volume}
  {50}},\ \bibinfo {pages} {6357} (\bibinfo {year} {1994})},\ \Eprint
  {https://arxiv.org/abs/astro-ph/9407016} {arXiv:astro-ph/9407016}
  \BibitemShut {NoStop}%
\bibitem [{\citenamefont {Espinosa}\ \emph {et~al.}(2008)\citenamefont
  {Espinosa}, \citenamefont {Giudice},\ and\ \citenamefont
  {Riotto}}]{Espinosa:2007qp}%
  \BibitemOpen
  \bibfield  {author} {\bibinfo {author} {\bibfnamefont {J.~R.}\ \bibnamefont
  {Espinosa}}, \bibinfo {author} {\bibfnamefont {G.~F.}\ \bibnamefont
  {Giudice}},\ and\ \bibinfo {author} {\bibfnamefont {A.}~\bibnamefont
  {Riotto}},\ }\href {https://doi.org/10.1088/1475-7516/2008/05/002} {\bibfield
   {journal} {\bibinfo  {journal} {JCAP}\ }\textbf {\bibinfo {volume} {05}},\
  \bibinfo {pages} {002}},\ \Eprint {https://arxiv.org/abs/0710.2484}
  {arXiv:0710.2484 [hep-ph]} \BibitemShut {NoStop}%
\bibitem [{\citenamefont {Hook}\ \emph {et~al.}(2015)\citenamefont {Hook},
  \citenamefont {Kearney}, \citenamefont {Shakya},\ and\ \citenamefont
  {Zurek}}]{Hook:2014uia}%
  \BibitemOpen
  \bibfield  {author} {\bibinfo {author} {\bibfnamefont {A.}~\bibnamefont
  {Hook}}, \bibinfo {author} {\bibfnamefont {J.}~\bibnamefont {Kearney}},
  \bibinfo {author} {\bibfnamefont {B.}~\bibnamefont {Shakya}},\ and\ \bibinfo
  {author} {\bibfnamefont {K.~M.}\ \bibnamefont {Zurek}},\ }\href
  {https://doi.org/10.1007/JHEP01(2015)061} {\bibfield  {journal} {\bibinfo
  {journal} {JHEP}\ }\textbf {\bibinfo {volume} {01}},\ \bibinfo {pages}
  {061}},\ \Eprint {https://arxiv.org/abs/1404.5953} {arXiv:1404.5953 [hep-ph]}
  \BibitemShut {NoStop}%
\bibitem [{\citenamefont {Kearney}\ \emph {et~al.}(2015)\citenamefont
  {Kearney}, \citenamefont {Yoo},\ and\ \citenamefont
  {Zurek}}]{Kearney:2015vba}%
  \BibitemOpen
  \bibfield  {author} {\bibinfo {author} {\bibfnamefont {J.}~\bibnamefont
  {Kearney}}, \bibinfo {author} {\bibfnamefont {H.}~\bibnamefont {Yoo}},\ and\
  \bibinfo {author} {\bibfnamefont {K.~M.}\ \bibnamefont {Zurek}},\ }\href
  {https://doi.org/10.1103/PhysRevD.91.123537} {\bibfield  {journal} {\bibinfo
  {journal} {Phys. Rev. D}\ }\textbf {\bibinfo {volume} {91}},\ \bibinfo
  {pages} {123537} (\bibinfo {year} {2015})},\ \Eprint
  {https://arxiv.org/abs/1503.05193} {arXiv:1503.05193 [hep-th]} \BibitemShut
  {NoStop}%
\bibitem [{\citenamefont {Espinosa}\ \emph {et~al.}(2015)\citenamefont
  {Espinosa}, \citenamefont {Giudice}, \citenamefont {Morgante}, \citenamefont
  {Riotto}, \citenamefont {Senatore}, \citenamefont {Strumia},\ and\
  \citenamefont {Tetradis}}]{Espinosa:2015qea}%
  \BibitemOpen
  \bibfield  {author} {\bibinfo {author} {\bibfnamefont {J.~R.}\ \bibnamefont
  {Espinosa}}, \bibinfo {author} {\bibfnamefont {G.~F.}\ \bibnamefont
  {Giudice}}, \bibinfo {author} {\bibfnamefont {E.}~\bibnamefont {Morgante}},
  \bibinfo {author} {\bibfnamefont {A.}~\bibnamefont {Riotto}}, \bibinfo
  {author} {\bibfnamefont {L.}~\bibnamefont {Senatore}}, \bibinfo {author}
  {\bibfnamefont {A.}~\bibnamefont {Strumia}},\ and\ \bibinfo {author}
  {\bibfnamefont {N.}~\bibnamefont {Tetradis}},\ }\href
  {https://doi.org/10.1007/JHEP09(2015)174} {\bibfield  {journal} {\bibinfo
  {journal} {JHEP}\ }\textbf {\bibinfo {volume} {09}},\ \bibinfo {pages}
  {174}},\ \Eprint {https://arxiv.org/abs/1505.04825} {arXiv:1505.04825
  [hep-ph]} \BibitemShut {NoStop}%
\bibitem [{\citenamefont {East}\ \emph {et~al.}(2017)\citenamefont {East},
  \citenamefont {Kearney}, \citenamefont {Shakya}, \citenamefont {Yoo},\ and\
  \citenamefont {Zurek}}]{East:2016anr}%
  \BibitemOpen
  \bibfield  {author} {\bibinfo {author} {\bibfnamefont {W.~E.}\ \bibnamefont
  {East}}, \bibinfo {author} {\bibfnamefont {J.}~\bibnamefont {Kearney}},
  \bibinfo {author} {\bibfnamefont {B.}~\bibnamefont {Shakya}}, \bibinfo
  {author} {\bibfnamefont {H.}~\bibnamefont {Yoo}},\ and\ \bibinfo {author}
  {\bibfnamefont {K.~M.}\ \bibnamefont {Zurek}},\ }\href
  {https://doi.org/10.1103/PhysRevD.95.023526} {\bibfield  {journal} {\bibinfo
  {journal} {Phys. Rev. D}\ }\textbf {\bibinfo {volume} {95}},\ \bibinfo
  {pages} {023526} (\bibinfo {year} {2017})},\ \Eprint
  {https://arxiv.org/abs/1607.00381} {arXiv:1607.00381 [hep-ph]} \BibitemShut
  {NoStop}%
\bibitem [{\citenamefont {Noorbala}\ \emph {et~al.}(2018)\citenamefont
  {Noorbala}, \citenamefont {Vennin}, \citenamefont {Assadullahi},
  \citenamefont {Firouzjahi},\ and\ \citenamefont {Wands}}]{Noorbala:2018zlv}%
  \BibitemOpen
  \bibfield  {author} {\bibinfo {author} {\bibfnamefont {M.}~\bibnamefont
  {Noorbala}}, \bibinfo {author} {\bibfnamefont {V.}~\bibnamefont {Vennin}},
  \bibinfo {author} {\bibfnamefont {H.}~\bibnamefont {Assadullahi}}, \bibinfo
  {author} {\bibfnamefont {H.}~\bibnamefont {Firouzjahi}},\ and\ \bibinfo
  {author} {\bibfnamefont {D.}~\bibnamefont {Wands}},\ }\href
  {https://doi.org/10.1088/1475-7516/2018/09/032} {\bibfield  {journal}
  {\bibinfo  {journal} {JCAP}\ }\textbf {\bibinfo {volume} {09}},\ \bibinfo
  {pages} {032}},\ \Eprint {https://arxiv.org/abs/1806.09634} {arXiv:1806.09634
  [hep-th]} \BibitemShut {NoStop}%
\bibitem [{\citenamefont {Fumagalli}\ \emph {et~al.}(2020)\citenamefont
  {Fumagalli}, \citenamefont {Renaux-Petel},\ and\ \citenamefont
  {Ronayne}}]{Fumagalli:2019ohr}%
  \BibitemOpen
  \bibfield  {author} {\bibinfo {author} {\bibfnamefont {J.}~\bibnamefont
  {Fumagalli}}, \bibinfo {author} {\bibfnamefont {S.}~\bibnamefont
  {Renaux-Petel}},\ and\ \bibinfo {author} {\bibfnamefont {J.~W.}\ \bibnamefont
  {Ronayne}},\ }\href {https://doi.org/10.1007/JHEP02(2020)142} {\bibfield
  {journal} {\bibinfo  {journal} {JHEP}\ }\textbf {\bibinfo {volume} {02}},\
  \bibinfo {pages} {142}},\ \Eprint {https://arxiv.org/abs/1910.13430}
  {arXiv:1910.13430 [hep-ph]} \BibitemShut {NoStop}%
\bibitem [{\citenamefont {Risken}\ and\ \citenamefont
  {Haken}(1989)}]{risken1989fpe}%
  \BibitemOpen
  \bibfield  {author} {\bibinfo {author} {\bibfnamefont {H.}~\bibnamefont
  {Risken}}\ and\ \bibinfo {author} {\bibfnamefont {H.}~\bibnamefont {Haken}},\
  }\href@noop {} {\emph {\bibinfo {title} {The Fokker-Planck Equation: Methods
  of Solution and Applications Second Edition}}}\ (\bibinfo  {publisher}
  {Springer},\ \bibinfo {year} {1989})\BibitemShut {NoStop}%
\bibitem [{\citenamefont {Motohashi}\ \emph {et~al.}(2012)\citenamefont
  {Motohashi}, \citenamefont {Suyama},\ and\ \citenamefont
  {Yokoyama}}]{Motohashi:2012bb}%
  \BibitemOpen
  \bibfield  {author} {\bibinfo {author} {\bibfnamefont {H.}~\bibnamefont
  {Motohashi}}, \bibinfo {author} {\bibfnamefont {T.}~\bibnamefont {Suyama}},\
  and\ \bibinfo {author} {\bibfnamefont {J.}~\bibnamefont {Yokoyama}},\ }\href
  {https://doi.org/10.1103/PhysRevD.86.123514} {\bibfield  {journal} {\bibinfo
  {journal} {Phys. Rev. D}\ }\textbf {\bibinfo {volume} {86}},\ \bibinfo
  {pages} {123514} (\bibinfo {year} {2012})},\ \Eprint
  {https://arxiv.org/abs/1210.2497} {arXiv:1210.2497 [hep-th]} \BibitemShut
  {NoStop}%
\bibitem [{\citenamefont {Markkanen}\ \emph {et~al.}(2019)\citenamefont
  {Markkanen}, \citenamefont {Rajantie}, \citenamefont {Stopyra},\ and\
  \citenamefont {Tenkanen}}]{Markkanen:2019kpv}%
  \BibitemOpen
  \bibfield  {author} {\bibinfo {author} {\bibfnamefont {T.}~\bibnamefont
  {Markkanen}}, \bibinfo {author} {\bibfnamefont {A.}~\bibnamefont {Rajantie}},
  \bibinfo {author} {\bibfnamefont {S.}~\bibnamefont {Stopyra}},\ and\ \bibinfo
  {author} {\bibfnamefont {T.}~\bibnamefont {Tenkanen}},\ }\href
  {https://doi.org/10.1088/1475-7516/2019/08/001} {\bibfield  {journal}
  {\bibinfo  {journal} {JCAP}\ }\textbf {\bibinfo {volume} {08}},\ \bibinfo
  {pages} {001}},\ \Eprint {https://arxiv.org/abs/1904.11917} {arXiv:1904.11917
  [gr-qc]} \BibitemShut {NoStop}%
\bibitem [{\citenamefont {Markkanen}\ and\ \citenamefont
  {Rajantie}(2020)}]{Markkanen:2020bfc}%
  \BibitemOpen
  \bibfield  {author} {\bibinfo {author} {\bibfnamefont {T.}~\bibnamefont
  {Markkanen}}\ and\ \bibinfo {author} {\bibfnamefont {A.}~\bibnamefont
  {Rajantie}},\ }\href {https://doi.org/10.1088/1475-7516/2020/03/049}
  {\bibfield  {journal} {\bibinfo  {journal} {JCAP}\ }\textbf {\bibinfo
  {volume} {03}},\ \bibinfo {pages} {049}},\ \Eprint
  {https://arxiv.org/abs/2001.04494} {arXiv:2001.04494 [gr-qc]} \BibitemShut
  {NoStop}%
\bibitem [{\citenamefont {Fernandez~C.}(2010)}]{SUSYQM}%
  \BibitemOpen
  \bibfield  {author} {\bibinfo {author} {\bibfnamefont {D.~J.}\ \bibnamefont
  {Fernandez~C.}},\ }\href {https://doi.org/10.1063/1.3507423} {\bibfield
  {journal} {\bibinfo  {journal} {AIP Conf. Proc.}\ }\textbf {\bibinfo {volume}
  {1287}},\ \bibinfo {pages} {3} (\bibinfo {year} {2010})},\ \Eprint
  {https://arxiv.org/abs/0910.0192} {arXiv:0910.0192 [quant-ph]} \BibitemShut
  {NoStop}%
\bibitem [{\citenamefont {Chialva}(2011)}]{Chialva:2010jt}%
  \BibitemOpen
  \bibfield  {author} {\bibinfo {author} {\bibfnamefont {D.}~\bibnamefont
  {Chialva}},\ }\href {https://doi.org/10.1103/PhysRevD.83.023512} {\bibfield
  {journal} {\bibinfo  {journal} {Phys. Rev. D}\ }\textbf {\bibinfo {volume}
  {83}},\ \bibinfo {pages} {023512} (\bibinfo {year} {2011})},\ \Eprint
  {https://arxiv.org/abs/1004.2051} {arXiv:1004.2051 [astro-ph.CO]}
  \BibitemShut {NoStop}%
\bibitem [{\citenamefont {Camargo-Molina}\ \emph {et~al.}(2022)\citenamefont
  {Camargo-Molina}, \citenamefont {Carrillo~Gonz\'alez},\ and\ \citenamefont
  {Rajantie}}]{Camargo-Molina:2022nnn}%
  \BibitemOpen
  \bibfield  {author} {\bibinfo {author} {\bibfnamefont {J.}~\bibnamefont
  {Camargo-Molina}}, \bibinfo {author} {\bibfnamefont {M.}~\bibnamefont
  {Carrillo~Gonz\'alez}},\ and\ \bibinfo {author} {\bibfnamefont
  {A.}~\bibnamefont {Rajantie}},\ }\href@noop {} {\  (\bibinfo {year}
  {2022})},\ \Eprint {https://arxiv.org/abs/2202.XXXX} {arXiv:2202.XXXX
  [hep-ph]} \BibitemShut {NoStop}%
\end{thebibliography}%

\end{document}